# On Modelling Life

Chris Adami

W.K. Kellogg Radiation Laboratory 106-38
California Institute of Technology
Pasadena, CA 91125

## Abstract

We present a theoretical as well as experimental investigation of a population of self-replicating segments of code subject to random mutation and survival of the fittest. Under the assumption that such a system constitutes a minimal system with characteristics of life, we obtain a number of statements on the evolution of complexity and the trade-off between entropy and information.

## 1 Introduction

What are the defining characteristics of life? For biologists and those studying living systems, this is the question that predates other questions and generates passionate arguments; for there are several ways how to answer this question. On the one hand, there is the philosophical approach. We would invent a "Turing Test of Life" in the tradition of the empiricists: "Anything that appears to be alive, is alive". On the other hand, we may draw up a list of characteristics and decide the question case by case, through a look-up table of sorts. To the scientist interested in the foundations of life, however, neither of these approaches is satisfactory. Rather, the scientific method applied to the problem of obtaining the defining characteristics of life would demand that we construct a *model* of life, one that displays the necessary attributes to pass the "Turing Test of Life", and then *strip it down to the most simple system that still displays these characteristics*. Anything that, if removed from this most simple system would result in a drastic change, removing the life-likeness, must be considered as a defining characteristic of this system. This does not preclude the existence of other systems passing the same test, with different characteristics. It does however establish the model under consideration as a baseline, and its characteristics as universal in its class. It is after this step has been taken that we can further enrich the model to learn more about life.

Clearly, this kind of approach is impossible using natural systems, as all life currently extant shows considerably more sophistication than needed just to survive. Thus, the construction of the minimally living system must be artificial. This has recently become possible with the advent of Tom Ray's tierra [1-3], and systems based on his idea. Briefly, tierra-like, (or "auto-adaptive genetic" [4]) systems consist of a population of self-replicating strings of information coded in a suitable manner that "live", and coevolve in an environment subject to mutation. The information contained in the self-replicating strings is the information necessary to carry out the task of self-replication, as well as any information that might give a certain genotype an advantage over another. Typically, this is information about the population itself, and the environment it lives in. Such a system displays a wealth of complexity and "life-like" characteristics [1-6]; enough to suggest answers to a number of fascinating questions relating to life.

Our approach then will be this: We assume that such systems of self-replicating strings *do* constitute a "minimal living system", and if not minimal then close enough to support the notion that the results obtained using the model are reflections of its "aliveness", rather than artifacts. While the theoretical results are independent of a particular implementation, the experimental ones are not. It is beyond the scope of this article to examine how close the particular system is to the abstract minimal living system, yet our results suggest that they are a good approximations, and in many respects universal. In the following we present theoretical arguments and experiments supporting them that suggest answers to a number of fascinating problems, including the emergence of complexity in living systems, and the nature of the evolutionary process.

Let us attempt to isolate the main characteristics of the "minimal living system". Primarily, we find that the cornerstone of this system is self-replication. The statistical mechanics of the ensemble of strings with and without self-replication are so vastly different (see Section 2) that we can determine with confidence to have isolated a defining characteristic of living systems. While this in itself is not surprising, it is specifically the self-replication of the information contained in the genome that is necessary (and that in this simple system is tantamount to replication of the entity). The consequences of this fact alone are manifold. Considering the evidence for genetic similarity in tRNA from the time of the origin of life to today [7], this capability seems to allow for the conservation of information over infinite periods of time in a noisy environment.

The keys for the evolution of complexity in this system are: bit-wise mutation acting on the genome of members of the population, and survival of the fittest. The former is again an obvious choice for a necessary condition: Without mutation, a population of self-replicating strings shows no patterns, no evolution, and no complexity. Clearly, mutation works against the force that attempts to preserve information, yet at the same time mutation *creates* information, or rather transfers it from the environment into the genome of the strings, where again it will be preserved by self-replication. We find that "survival of the fittest" is synonymous with *survival of the most populous*. Thus, fitness in this system is directly related to replication rate. In the absence of a mechanism that enables one string to prevent the replication of another (kill-strategy), it is conceivable that "fitness = replication-rate" is a universal feature of such low-level organisms.

Another mystery of living systems is associated with the evolution of complexity and its apparent contradiction with the second law of thermodynamics, the law of increasing entropy. While it is known that dissipative systems involving microscopic irreversibility can evolve from disorder to order, the precise mechanism is often unclear. In this model, the evolution of complexity appears to proceed by a trade-off between entropy and information. Information is transferred from the environment into the genome of the population via chance mutations that allow the mutated string to exploit the information for a higher replication rate. As a consequence, the (information-theoretic) entropy of the population of strings drops by a commensurate amount until off-spring of the mutated string have reestablished disorder and the previous level of entropy is (roughly) restored. The time between such events is distributed according to a power law, and we thus observe periods of stasis interrupted by "avalanches of invention". Again, if we accept this system as a model for the most primitive and basic living system, the evidence suggests that evolution is Darwininan in nature even though not gradual, but rather punctuated: the picture promoted by Gould and Eldredge [8]. Below, we shall give more details and results of experiments that support the claims made here.

## 2 Statistical Mechanics of Self-Replicating Bit-Strings

Let us approach the model system from a theoretical point of view and write down a few of the basic equations governing the dynamics of self-replicating strings. Consider a system of $N$ strings of $\ell$ instructions, where each instruction can take on $p$ values ($p = 2$ for binary strings, $p = 4$ for DNA, and $p = 32$ for Ray's tierra). Also, let there be $N_g$ different genotypes (different combinations of instructions) extant in the population, $N_g \leq N$, and let $n_i$ denote the number of strings of genotype $i$. Then,

$$N = \sum_i^{N_g} n_i \ . \tag{1}$$

Let us further assume that the total number of strings in the population is constant: $dN/dt = 0$ (this constraint can be relaxed trivially). Also, denote the replication rate (off-spring per unit time) of genotype $i$ by $\varepsilon_i$. Defining the average replication rate (average "fitness")

$$\bar{\varepsilon} = \sum_i^{N_g} \frac{n_i}{N} \varepsilon_i \equiv \sum_i^{N_g} \rho_i \varepsilon_i \ , \tag{2}$$

where $\rho_i$ is the genotype *density* of species $i$, we can immediately write down the equation that describes the time-evolution of $n_i$ in a "mean-field" approximation

$$\begin{aligned} n_i(t+1) - n_i(t) &= (\varepsilon_i - \langle \varepsilon \rangle - R\ell) \ n_i(t) + \frac{N}{N_g} R\ell \\ &\equiv \gamma_i n_i(t) + C \end{aligned} \tag{3}$$

with obvious definitions for $\gamma_i$ and $C$, the "source" term. Here, we introduced the mutation rate $R$ in units [mutations·(site·time)$^{-1}$]. Thus, $R\ell$ is the probability that a string of length $\ell$ is hit by a mutation per unit time, i.e. per instruction executed. The constant term on the right hand side of (3) is modelling mutations acting on genotypes $j \neq i$ and ensures $\dot{N} = 0$. Equation (3) can be solved in the equilibrium limit, $d\bar{\varepsilon}/dt = 0$:

$$n_i(t) = \frac{N}{N_g} \left\{ (1 + \frac{R\ell}{\gamma_i}) e^{\gamma_i t} - \frac{R\ell}{\gamma_i} \right\} \tag{4}$$

using an equilibrium boundary condition, $n_i(0) = N/N_g = \bar{n}$. For $C \approx 0$ this is simply

$$n_i(t) = n_i(0) e^{\gamma_i t} \tag{5}$$

where $\gamma_i = \varepsilon_i - \bar{\varepsilon} - R\ell$ is the "growth factor". Clearly, in equilibrium $\gamma_i = 0$ only for the best genotype and $\gamma_i < 0$ for inferior ones.

In order to make contact with statistical mechanics, define the *inferiority*

$$E_i = \varepsilon_{\text{best}} - \varepsilon_i \tag{6}$$

where $\varepsilon_{\text{best}}$ is the highest replication rate in the population. Then, $E_i \geq 0$ and we can define the ground-state ("vacuum") of the population by

$$\langle 0|E|0 \rangle = \bar{E} = 0 \ . \tag{7}$$

Let us also define the "fine-grained" (Shannon-) entropy

$$S = -\sum_i^{N_g} \frac{n_i}{N} \log(\frac{n_i}{N}) \tag{8}$$

with the property

$$\frac{\partial S}{\partial E} \simeq \frac{1}{R\ell} \ , \tag{9}$$

which suggests that the system of self-replicating strings is analogous to atoms with eigen-energies $E_i$ in a heat bath of temperature $R\ell$ which initiates transitions $E_i \to E_j$ via mutations. In fact, detailed balance arguments lead to the equation determining the equilibrium distribution $n_i(E_i, \bar{E})$

$$\frac{1}{n_i} - \frac{1}{n_j} = \frac{1}{C}(E_i - E_j) \tag{10}$$

with the solution

$$n_i(E_i, \bar{E}) = \frac{C}{E_i - \bar{E} + R\ell}. \quad (11)$$

In equilibrium, we find $\bar{E} \to R\ell$, universally. However, this equilibrium is disrupted by spontaneous phase transitions triggered by mutations creating genotypes with $\varepsilon_i > \varepsilon_{\text{best}}$ with a "latent heat" $L = \Delta\varepsilon = \varepsilon_{\text{best}}^{\text{new}} - \varepsilon_{\text{best}}^{\text{old}}$. After such an event the entropy of the system defined by (2) drops, and then rebounds. In Fig. 1 we compare the entropy of a system of strings subject to mutation with or without replication in the avida environment [6][1]. If we start both systems in the uniform state $\bar{E} = 0$ (which implies $S = 0$: all strings are identical), we see that the self-replicating population reaches a plateau with $S < \log(N)$ while the non-self-replicating system reaches the maximum value rather quickly. In addition, the population of self-replicating strings can lower the entropy spontaneously, a feat that is impossible for the non-replicating population which must strictly adhere to the second law of thermodynamics.

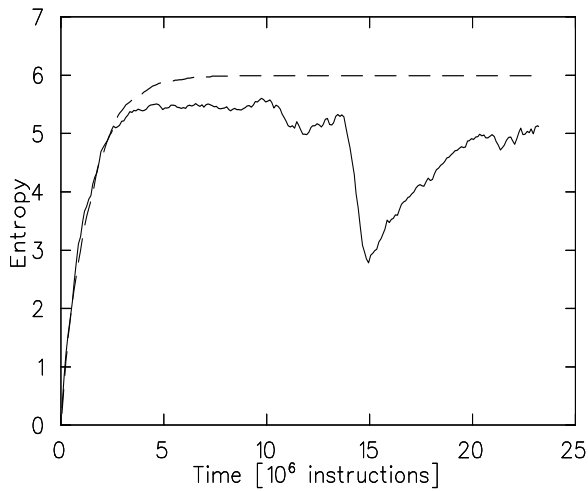

FIG. 1 Entropy of a population of 400 self-replicating (solid line) and non-replicating (dashed line) strings started in the uniform state.

## 3 Entropy and Information

Let us investigate more closely the time-evolution of entropy in systems of self-replicating entities. From the point of view of statistical mechanics rather than information theory, entropy is just the logarithm of the number of states in the volume of "phase-space" occupied by the population. Genetic "phase-space", or genotype-space, is spanned by the combinatorics to assemble the basic building blocks (or bits), and is finite for finite length strings. The population forms a "cloud" in genotype-space that moves slowly according to the fitness-landscape. Each genotype moves along a trajectory determined by a gradient term (adaptive process) and a noise term (mutations). As a consequence, the time-evolution of the genotype of a string is determined by the usual Langevin equation. While genotype-space is very high-dimensional, the number of states occupied by the population is by definition $N_g$, and thus,

$$S = \log(N_g) \quad (12)$$

which is equivalent to the fine-grained entropy in the large $N$ limit. The number of genotypes $N_g$ is an equilibrium quantity, depending on mutation rate $R\ell$, total number of strings $N$, and most importantly, the average inferiority of the population. Clearly, only in infinite systems after infinite time is $\log(N_g) = \ell\log(p)$. For finite systems, the number of genotypes is determined by the rate of birth of new genotypes and the rate of extinction of existing ones. The rate of birth of new, viable, genotypes is determined by the ratio of "cold" to "hot" spots in the genome. A cold spot on the string is an instruction that, if mutated, leads to either death or severe inferiority. Hot spots are instructions that are neutral under mutation and are thus responsible for the $\varepsilon$-degeneracy of genotypes. Thus, if there were no hot spots in the genome, there could not be any diversity, and in fact no evolution could take place, as was conjectured some time ago [9].

Let us denote the number of cold spots in a string of length $\ell$ by $\nu_c$. Then the birth rate of new genotypes is $R\ell N$, while the birth rate of *viable* genotypes is $(1 - \nu_c/\ell)R\ell N$. On the other hand, the extinction rate of existing (viable) genotypes depends (again in a mean-field approximation) on the average inferiority of the population, $\bar{E}$. Clearly, for zero inferiority there will be no extinctions, while extinctions are maximal during a phase transition, when the average inferiority is high. We thus find

$$\dot{N}_g = NR\ell(1 - \frac{\nu_c}{\ell}) - \bar{E}N_g \quad (13)$$

and thus in equilibrium

$$N_g = N\frac{R\ell}{\bar{E}}(1 - \frac{\nu_c}{\ell}). \quad (14)$$

We are now ready to examine the dropping of entropy observed through phase transitions. The decrease seen in Fig. 1 is of course due to the discovery of information and a corresponding new $\varepsilon_{\text{best}}$. Through such a transition, $\bar{E} \to \bar{E} + \Delta\varepsilon$ and $\Delta S = \Delta\varepsilon/(R\ell)$. However, with time $\bar{E}$ relaxes, and the entropy rebounds, but not necessarily to its previous equilibrium value. Consider for instance the situation where the information gained in the process requires more cold spots $\nu_c$ for storage than before. Then, a phase transition effectively freezes some spots that were originally hot spots, and thus reduces the effective length of the string. This is the situation in Fig. 1, the entropy rebounds, but reaches a lower plateau than before, due to the incorporation of information in the genome. We are thus witnessing a curious trade-off between information and entropy. In fact, it has been conjectured [10] that the physical entropy is not given by just the fine-grained entropy, but rather is the sum of Shannon entropy $S(\rho)$ and algorithmic complexity $K(\rho)$, the latter being a measure of information stored in a string. In this scenario, complexity can emerge without violating the second law, simply by trading information for entropy while keeping the physical entropy constant.

---

[1] We chose this system because the number of strings can be kept strictly constant in avida.

## 4 Fractal Structure of Evolution

Earlier we described how macroscopically the system is in an equilibrium state only rarely disrupted by chance mutations that happen to improve the fitness of the population. Here, we would like to take a more quantitative look at this behaviour, and work out the consequences for the structure of evolution.

More and more evidence is surfacing that certain quantities relating to living systems are distributed according to power laws, such as the frequency of extinction events of a certain size [11]. It has been known for a long time that there is a structure in the taxonomic system, since it was pointed out by Willis [12] that there is a very large number of taxa with only one or a few subtaxa, but only a few taxa with many subtaxa. Recently, there has been an extensive analysis of the frequency distribution of the number of taxa $N$ that have $n$ sub-taxa [13], with the result that $N \sim n^{-D}$ with $D \approx 2 \pm 0.5$. This fractal structure of evolution has been obtained from a pain-staking analysis of fossil records from the Cambrian to the Tertiary (families within orders) as well as a selection of catalogued flora and fauna (subclades within clades, families within orders, and species within genera). Yet, an explanation for this structure could not be offered. Generally speaking, such a distribution can be expected if no scale determines the length of time that a certain species dominates a population, and the number of subfamilies is proportional to this period of time. This hypothesis can be tested using the minimal living system.

Recently [5], we suggested that self-replicating systems typically operate in a self-organized critical state [14]. Drawing the analogy to the sandpile model, we could identify information as the agent that causes the self-organization, and which is transported via a dissipative process through the system. We also identified the growth-factor $\gamma_i$ [see Eq. (5)] as the critical variable. In the equilibrium situation, the population is dominated by the current "best" genotype and its $\varepsilon$-degenerate off-spring, all of which have $\gamma_i = 0$. An advantageous mutation however can disrupt this equilibrium by creating a new "best" genotype with $\gamma_i > 0$. Thus, the vacuum with $\bar{E} = 0$ is now a false one, and a phase transition must occur. This happens in the manner of the sandpile avalanches, with the new information transmitted through the system via the off-spring. Gradually, all genotypes with a subcritical replication rate will become extinct and replaced: the system has returned to its critical state. A necessary condition for the identification of the self-organized critical state is a power spectrum of the fitness history (replication rate as a function of time) that is of the power law type.

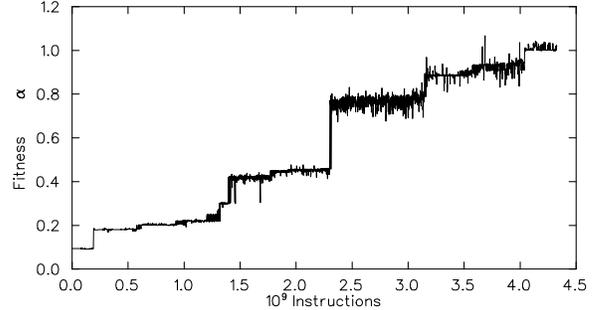

FIG. 2 Fitness curve for a typical run. The (normalized) fitness $\alpha$ of the most successful (i.e., most populous) genotype is plotted as a function of time, measured every million instructions for a mutation rate $R = 0.65 \times 10^{-3}$.

A typical fitness history (normalized replication rate of the most abundant genotype in the population as a function of evolutionary time) obtained using the tierra system is displayed in Fig. 2.

The associated power spectral density in Fig. 3 shows a near perfect $1/f^2$ spectrum over 4 orders of magnitude. While a power-law spectral density is a necessary condition for the identification of a self-organized critical state, it is not sufficient. Indeed, a $1/f^2$ spectrum is typical for a simple random walk, and thus for the most basic dissipative process. Incidentally, the process of evolution can be described as a guided random walk in genotype space, albeit a *non-Markovian* one.

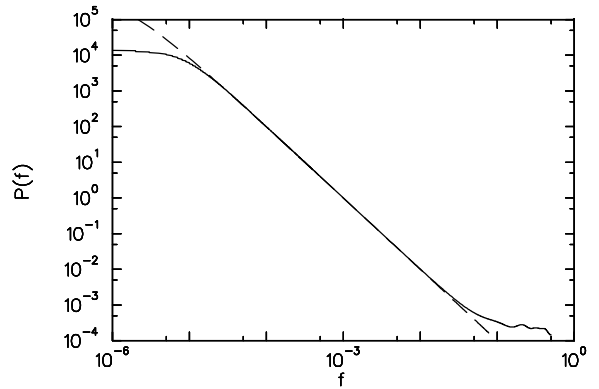

FIG. 3 Power spectrum $P(f)$ of a typical fitness curve $\alpha(t)$ (Fig. 2). The dashed line is a fit to $P(f) \sim f^{-\beta}$ with $\beta = 2.0 \pm 0.05$.

On the other hand, the distribution of waiting times between events of a certain size in a random walk model is exponential, rather than of the power law type. The latter is expected if the time between such events is not controlled by any scale. The fundamental time scales in self-replicating systems of the sort discussed here are the gestation time (the time for a string to gestate one off-spring) and the average time between two mutations hitting a single string (see Ref. [4]). Both of these time scales, however, are much smaller than the lifetime of single genotypes. Also, as discussed earlier in connection with the heat-bath analogy, the system can be understood as a superposition of a very large number of meta-stable states, with transitions between them induced by the mutations. In such a system, the time between avalanches or "events" is distributed according to a power law, as is the size and duration of the events [14]. Using the tierra system at a mutation rate of $R = 0.65 \times 10^{-3}$ (mutations per instruction executed), with a total number of 131,072 sites inhabited by typically 600-1,400 strings of length 60-150, we have measured the frequency distribution $N(\tau)$ of waiting times $\tau$ (lengths of "epochs") between phase transitions (which are signalled by a discontinuous jump in fitness as in Fig. 2). Figure 4 shows the *integrated* distribution function

$$M(\tau) = \frac{1}{\tau} \int_\tau^\infty N(t) dt \ , \qquad (15)$$

The integrated quantity is distributed with the same exponent as $N(t)$ (as can be shown by direct differentiation), but is more reliable due to improved statistics.

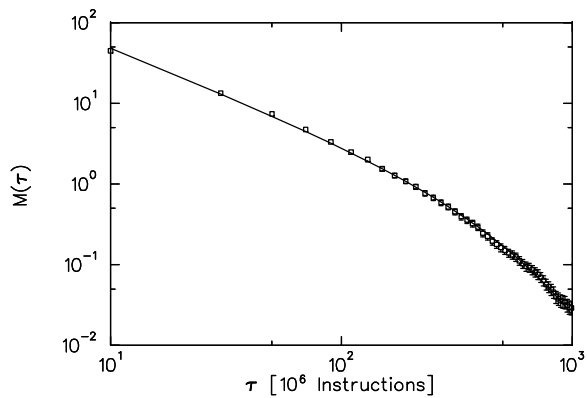

FIG. 4 Integrated distribution $M(\tau)$ of times between phase transitions $\tau$ (length of epoch) measured in millions of executed instructions.

The data were obtained from 50 runs under identical conditions (save the random number seed) resulting in 512 waiting times. The resulting curve was fit to the standard power law, cut-off by finite-size effects [2]

$$M(\tau) = c\ \tau^{-\alpha} \exp\left(-\frac{\tau}{T}\right) \qquad (16)$$

and obtained $\alpha = 1.1 \pm 0.05$ and $T = 450 \pm 50$, where all times are in units of millions of executed instructions. We expect the exponent of the power law to be universal, while the cut-off of course is not.

There are a number of fascinating consequences of such a distribution. If the time between phase transitions, as well as the size of fitness increases during such transitions are distributed according to a power law, curves such as the one depicted in Fig. 2 are fractal in nature, i.e. they (ideally) appear similar *at all scales* ("Devil's Staircase"[3]). If this is the case, we can imagine that a more sophisticated model, one that can run many orders of magnitudes longer than the one we used, with unlimited fitness increases, would yield just such a curve, where each element of the curve when examined closely would have the same structure as the one in Fig. 2. Then, we must admit that fitness increases of enormous magnitude can happen almost instantly, driven by microscopic point-mutations *only* [4], albeit extremely infrequently. Also, the model implies that at each such phase transition a large number of genotypes goes extinct dynamically [see Eq. (14)] without external interference (this was also observed in [3]). Thus, Raup's distribution of extinction events would equally follow.

On the other hand, it is tempting to believe that the longer a species dominates a population, the more subspecies will be generated. As a consequence, the frequency distribution of taxa with $n$ subtaxa, $N(n)$, must show the same critical exponent as the waiting time distribution [5].

## 5 Conclusions

We proposed to model life using artificial living systems that are "minimal" in their characteristics. We determined self-replication of information, bit-wise mutation of the information (providing the noise), and an environment of information (providing the gradient of the fitness-landscape) as necessary characters of such a model. The artificial life systems tierra and avida were used to determine universal characteristics of living systems, such as the evolution of complexity through transfer of information present in the environment into the genome, and its relation to the second law of thermodynamics. Furthermore, we were able to conjecture that living systems typically evolve towards a self-organized critical state and that this state determines fractal population structures. This also constitutes strong evidence for a punctuated equilibrium picture of evolution as favoured by the fossil record.

### Acknowledgments

This work was supported in part by NSF grant # PHY90-13248 and a Caltech Divisional fellowship.

---

[2] Since all runs were terminated after between 500 million and 4 billion instructions were executed, we cannot expect to measure the distribution of waiting-times larger than 500 million instructions with satisfactory accuracy.

[3] See, e.g., [15].

[4] In all of the previous discussion we have ignored mutations of the cross-over type, which effectively happen in systems such as tierra, but which can be viewed as *consequences* of point mutations. Very little is as yet known about the impact of cross-over mutations in the self-replicating systems described here, but we must assume that some of the phase transitions are in fact fitness improvements due to cross-over mutations.

[5] The discrepancy between Burlando's coefficient $D \approx 2$ and our coefficient $\alpha = 1$ is not understood.